# Graphene plasmonics: Physics and potential applications

*Shenyang Huang[1,2], Chaoyu Song[1,2], Guowei Zhang[1,2], and Hugen Yan[1,2]\**

[1]Department of Physics, State Key Laboratory of Surface Physics and Key Laboratory of Micro and Nano Photonic Structures (Ministry of Education), Fudan University, Shanghai 200433, China

[2]Collaborative Innovation Center of Advanced Microstructures, Nanjing 210093, China

*Email: hgyan@fudan.edu.cn

## Abstract

Plasmon in graphene possesses many unique properties. It originates from the collective motion of massless Dirac fermions and the carrier density dependence is distinctively different from conventional plasmons. In addition, graphene plasmon is highly tunable and shows strong energy confinement capability. Most intriguing, as an atom-thin layer, graphene and its plasmon are very sensitive to the immediate environment. Graphene plasmons strongly couple to polar phonons of the substrate, molecular vibrations of the adsorbates, and lattice vibrations of other atomically thin layers. In this review paper, we'll present the most important advances in grapene plasmonics field. The topics include terahertz plasmons, mid-infrared plasmons, plasmon-phonon interactions and potential applications. Graphene plasmonics opens an avenue for reconfigurable metamaterials and metasurfaces. It's an exciting and promising new subject in the nanophotonics and plasmonics research field.

1 Introduction

Graphene is a fascinating electronic and optical material and the study for graphene started with its magneto-transport measurements and an anomalous Berry phase[1, 2]. Optical properties of graphene attracted more and more attention later on. Many exciting developments have been achieved, such as universal optical conductivity[3], tunable optical absorption[4, 5] and strong terahertz response[6]. Most importantly, as an interdisciplinary topic, plasmon in graphene has been the major focus of graphene photonics in recent years. Both plasmonics community and graphene community are quite interested in graphene plasmonics. Due to the disadvantageous factors of metal-based plasmons[7], such as high loss and lack of tunability, the plasmonics community has been actively looking for new plasmonic materials[8]. Many different materials, ranging from semiconductors to conductive oxides and dielectric materials have been examined in terms of their plasmonic properties. People expect big breakthrough in such endeavor.

However, only limited success has been achieved. The advent of graphene opens new avenues for plasmonics research. Typically, there are three important merits for graphene as a plasmonic material, which can complement noble metals in plasmonics. Firstly, graphene has high carrier mobility, which indicates a relatively low loss as a plasmonic material. Secondly, graphene is a tunable material and the Fermi level of graphene can be readily tuned through electrostatic gating or chemical doping[5]. This guarantees that graphene can be an important building block for active plasmonic materials and metamaterials[9]. Thirdly, graphene plasmon shows very strong light confinement. The wavelength of plasmon is typically two orders of magnitude smaller than the light wavelength[10].

In the physics community, people are more interested in the fundamental properties of Dirac plasmons in graphene. Many-body interaction is an intriguing topic for physicists. Plasmon, the collective oscillation of charge carriers, is inherently a many-body subject. Before optical studies of graphene plasmons, EELS(electron energy loss spectroscopy) measurements studied the high energy plasmons($\pi$ plasmon and $\sigma$ plasmon) [11]. Optical studies of graphene plasmon in the early stage were mostly motivated by terahertz plasmon studies in traditional two-dimensional (2D) electron gas systems started from the 1970s[12, 13]. Exotic behaviors were revealed as a Dirac plasmon system[14, 15]. Due to the large mismatch between the plasmon momentum and photon momentum, graphene plasmons cannot be directly excited by light. Three major techniques have been utilized. The first one is to excite localized plasmons in graphene micro- and nano-structures[14-19]. The second one is to utilize a metallic tip to excite a propagating plasmon wave[20, 21]. This technique is based on the scattering-type scanning near field optical microscopy(s-SNOM) setup. The third method to couple light is through the use of gratings or prisms[7, 22]. The first two schemes have been widely used and the third one has only limited demonstrations.

Many research groups have been involved in graphene palsmonics and a large number of exciting results have been published. Due to the one-atom thick nature of graphene, graphene plasmons are extremely sensitive to the environment. There are multiple studies concerning the plasmon-substrate phonon coupling[16, 17, 23], plasmon-molecular vibration coupling[24, 25] and the coupling between graphene plasmons and phonon polaritons in atomic-layer hexagonal boron nitride (h-BN) [26, 27]. Graphene plasmon is not only sensitive to the immediate environment, it can also be tuned by external fields, such as magnetic field, which has been demonstrated by Yan *et al.*[28] and Poumirol *et al.*[29] and others[30-32]. Plasmon damping mechanisms are important for any plasmonic material. Multiple damping mechanisms for graphene plasmon have been carefully analyzed and possible ways to reduce plasmon damping have been demonstrated or proposed[16, 28]. For instance, Woessner *et al.* demonstrated the large improvement of plasmon damping for graphene/h-BN heterostructures[33]. In terms of the plasmon frequency range, there have been efforts to push the resonance wavelength all the way from the terahertz to the near-infrared[34]. In addition, graphene plasmonic crystal in the terahertz frequency range has been demonstrated[35, 36], which shows the importance of the periodicity in the crystal structure.

It's difficult to cover all of the topics mentioned above and many more not mentioned. In addition, exciting developments for plasmon research in related materials systems, such as the surface of topological insulators[37] will not be covered either. There are a few excellent review papers in recent years for graphene plasmonics, which provide us a good overview of the field[38-43].

This review paper is organized as following. After the introduction part, we'll focus on graphene plasmons in the terahertz frequency range, the topics such as Dirac plasmon, magneto-plasmon will be touched on. Then graphene mid-infrared (mid-IR) plasmons will be discussed, both localized plasmons and propagating plasmons will be reviewed. The s-SNOM scheme to excite mid-IR plasmons in graphene is one of the focuses in the section. The plasmon-phonon coupling is the next topic and potential applications will be another section in the review. Finally, we'll present a summary and an outlook for the research field.

## 2 Graphene terahertz plasmons

The first optical study of graphene plasmons is in the terahertz frequency range[14]. More than 30 years before that, terahertz and far infrared measurements of terahertz plasmons in traditional 2D electron gas were extensively conducted[12, 13, 44]. It's very natural to extend such study to a relatively new 2D electron gas system, i.e., graphene. However, this is not a simple replica of previous effort. Graphene terahertz plasmon gives us a plethora of surprises.

In a broad frequency range from mid-IR to the visible, single layer graphene absorbs 2.3% of the incident light, which is a manifestation of the renowned universal optical conductivity[3, 45]. In the terahertz frequency range, however, the intraband optical transition can give much stronger optical response[6, 46]. A moderately doped graphene can absorb 40% of the incident radiation. Such strong light-matter interaction renders graphene a wide range of potential applications in the terahertz range, such as terahertz metamaterials[9, 14] and EMI(electromagnetic interference) shielding[15, 47].

Many materials can support plasmon polaritons and lots of the plasmon properties are governed by electrodynamics, without knowing the details of the underlying material[7]. For graphene, the closest counterpart is the traditional 2D electron gas. As a general property of 2D plasmon, the energy dispersion has a $\sqrt{q}$ dependence. This is true for any 2D plasmons and is a direct consequence of the particle conservation in a 2D system[48]. Due to the linear band structure of graphene, carriers are relativistic Dirac fermions with zero rest mass[2]. This has lots of consequences and makes graphene plasmon distinctively different from the traditional 2D electron gas.

Experimentally, optical measurements of graphene terahertz plasmons utilized microstructures, such as graphene micro-ribbons[14](Figure1A) and micro-disks[15, 47](Figure1D). In Figure1B, Ju *et al.* showed the ribbon width dependence of the plasmon resonance. A narrower ribbon array shows a higher resonance frequency, which is a result of the 2D plasmon dispersion. For a ribbon with width $W$, the first order dipolar plasmon mode approximately corresponds to a propagating plasmon wave with a wave-vector $q=\pi/W$ [49]. Since the plasmon dispersion follows $\sqrt{q}$ relation, the plasmon resonance frequency for graphene ribbons follows $\frac{1}{\sqrt{W}}$ scaling, which has been demonstrated by Ju *et al.* [14]. This is the same as plasmons in a traditional 2D electron gas.

Figure1C presents the plasmon frequency dependence on the Fermi level. It shows that the frequency is proportional to $\sqrt{E_f}$, where $E_f$ is the Fermi energy. Since in graphene, $E_f$ is propertional to $\sqrt{n}$, where $n$ is the carrier density, the frequency is proportional to $n^{1/4}$. This is distinctively different from the traditional 2D gas, which exhibits a $\sqrt{n}$ scaling relation. Carriers in graphene don't have a well defined effective mass, therefore the plasmon effective mass $m_p$ exhibits a dependence on the carrier density. Simple argument gives $m_p=E_f/v_f^2$, which is proportional to $\sqrt{n}$. In the formula, $v_f$ is the Fermi velocity. Since the plasmon frequency is inversely proportional to $\sqrt{m_p}$, it gives rise to a weaker $n$ dependence for the graphene plasmon frequency, $\omega_p \sim n^{1/4}$[15]. Compared to noble metal plasmons, graphene plasmon has tunability due to the tunable carrier density, which is promising for reconfigurable metamaterials. However, the tunability is not as strong due to the nature of Dirac plasmon. To more efficiently tune the plasmon frequency, adding more graphene layers to form a multilayer structure is a feasible way, as shown by Yan *et al.*[15, 47]. Figure1E displays the plasmon resonance spectra for graphene disk arrays with 1, 2 and 5 graphene layers. Both the plasmon frequency and resonance amplitude are dramatically enhanced. Figure1F shows the plasmon frequency dependence on the effective carrier density in the multilayer structure, which follows a $\sqrt{n}$ dependence, rather than a $n^{1/4}$ dependence. For comparison, a data set for a single layer disk is also shown, which exhibits a much weaker $n^{1/4}$ dependence. This can be understood from the plasmon mass point of view[50]. For a single layer, when one increase the carrier density, the plasmon mass also increases, which slows down the increase of plasmon frequency. For the case of stacked multiple graphene layers, one can effectively increase the carrier density of the thin disk, without increasing the plasmon mass.

Graphene plasmon mass is tunable and has the same form as cyclotron mass in a magnetic field. By combining cyclotron resonance and plasmon resonance, Yan *et al.* have demonstrated novel magneto-plasmon behavior of graphene[28]. When graphene micro-disk array is placed in a magnetic field with the field perpendicular to the disk(Figure2A), the plasmon peak splits into two modes, one is a bulk mode and the other edge mode. Figure2B shows the typical magnetoplasmon spectrum and two distinct modes are present. Figure2A illustrates the carrier trajectory for those two modes. The edge modes features a rotating current around the disk edge and the period for a complete trip around the disk can give the edge plasmon frequency. The bulk plasmon mode features charge carriers inside the disk which undergo collective cyclotron resonance in the large field limit. It's worth noting that the linewidths for the two modes are very different, the edge mode is very sharp while the bulk mode is broadened. Figure2C plots the FWHM(full width at half maximum) for both modes as a function of the magnetic field. Solid curves are based on a classical model which starts from the equation of motion of individual charge carriers in a magnetic field. We see that the edge mode becomes sharper and sharper with an increasing magnetic field. This is because the edge mode is a current around the edge and it's one-dimensional nature can suppress carrier back-scattering in a magnetic field. Therefore, the carriers can travel around the edge for a long time and the plasmon mode has a long lifetime. The skipping orbits for the edge mode in Figure2A illustrates the suppression of the back scattering. Thanks to the long lifetime , the edge plasmon mode can be studied in the time domain. Petković *et al.*[32] and Kumada *et al.*[31] monitored the pulse current around the edge of a large area graphene piece, with size up to the millimeter. Figure2D sketches the experimental configuration, a pulse current is injected in one point and can be detected in another point of the graphene edge in a later time. Switching the direction of the magnetic field can switch the current direction, which usually gives a different delay time in the detection point. This is the so called chiral property[32]. Figure2E shows the velocity of the edge mode carriers at different magnetic field. The velocity can vary quite dramatically. Other factors, such as carrier density and screening can also affect the carrier drift velocity[32].

Plasmonic coupling between graphene microstructures have been investigated by Yan *et al.*[19] and Rana *et al.*[51] Moreover, the periodical anti-dot structure of graphene have been carefully studied by Yeung *et al.*[35, 52] and Liu *et al.*[36] in terahertz frequencies. In those structures, graphene plasmon can form certain bands and multiple resonance peaks can be observed. All these studies go beyond individual microstructures and give more complicated behavior.

## 3 Graphene mid-infrared plasmon

For localized plasmons in graphene, when one shrink the size of the graphene structure from micron to 100-nanometer size, the plasmon frequency increases from terahertz to the mid-IR [16-18, 23]. In the mid-IR frequency range, the following can be expected. First, since the structure is small, the edges play a significant role in plasmon damping; Second, many of the lattice and molecule vibration frequencies are in the mid-IR range and an even stronger coupling to the environment is expected; Third, higher energy plasmon has more interactions with intrinsic phonons in graphene , which is believed to affect the plasmon lifetime.

Since 2013, multiple research groups have published results on localized mid-IR plasmons in graphene[16-18, 23, 24, 53]. Yan *et al.* studied graphene nanoribbons on both polar and non-polar substrates[16]. The damping pathways, which include edge scattering and optical phonon emission, were clearly identified. Figure3A shows the SEM image of graphene nanoribbons on $SiO_2$/Si substrate and the extinction spectra in the mid-IR range are presented in Figure3B, with incident light polarization perpendicular and parallel to the ribbons. Plasmons can be excited with perpendicular light polarization and multiple peaks are prominent in the plot. The three major peaks are hybrid modes involving one plasmon mode of graphene and two phonon-polariton modes of $SiO_2$. Some other examples of plasmon-phonon hybridization will be presented in the next section. Yan *et al.*[16] calculated the loss function of graphene plasmon as a function of the frequency and wave-vector through RPA (random phase approximation)method[54], by taking into account the plasmon-polar phonon interaction. Figure3C displays the calculated results and three branches of the hybrid modes are evident. In the figure, the experimentally measured plasmon frequencies are also indicated, with a good agreement. Later on, Luxmoore *et al.* simultaneously observed four hybrid modes in a broader frequency range for graphene nanoribbons on $SiO_2$, which showed a more complete picture[55].

The interaction with surface polar phonons is not the only story. When the plasmon frequency is higher than the intrinsic optical phonon frequency(~1580$cm^{-1}$) of graphene, a new plasmon decay channel is opened up. Figure3D plots the plasmon resonance spectra with resonance frequencies spanning from below to above the phonon frequency. It's evident that the linewidth increases abruptly, as shown in the inset of the figure. This is a clear indication of additional damping channel. Theory predicts that when the plasmon frequency is higher than the optical phonon frequency, it can enter into the intraband Landau damping regime by emitting an optical phonon[56]. This mechanism is sketched in Figure3E. It's an unfortunate fact because this will largely limit the plasmon quality factor when one pushes graphene plasmon to higher and higher frequencies[34].

In addition to localized mid-IR plasmons, the propagating plasmons in graphene have been intensively investigated by Fei *et al.* [20, 57]and Chen *et al.*[21] and other groups[58, 59]. The experimental technique is based on infrared nano-imaging by utilizing a s-SNOM. Many exciting results have been obtained, which include direct imaging of plasmon reflection from grain boundaries[57], identify long lifetime plasmons for graphene on h-BN substrates[33], and ultrafast pump-probe measurements for plasmon dynamics[60].

Figure4A is an artistic representation of the s-SNOM studies of graphene plasmon[20]. A sharp metal AFM tip is the key to excite propagating plasmons in the system, which can compensate the large momentum mismatch between plasmons and photons.  When the propagating plasmon meets edges or grain boundaries ,it can form standing waves between the tip and the edges or grain boundaries. Figure4B shows a typical imaging for a triangle shape graphene on SiO$_2$/Si substrate. Bright and dark fringes are prominent, from which the plasmon wavelength can be inferred. Graphene plasmons are highly tunable and the wavelength can be readily tuned by a gate voltage. Figure4C plots the gate dependence of the imaging line profile along a trace perpendicular to an edge of graphene. The profile has strong dependence on the gate voltage. The plasmon wavelength can be extracted from the data in Figure4C. Figure4D shows the gate dependence of the plasmon wavelength and the inset shows the extracted optical conductivity from the imaging data. Both of the real and imaginary parts of the optical conductivity exhibit gate dependence. Most importantly, graphene plasmon can be switched on/off with a gate voltage[21]. Upper panel of Figure4E illustrates the switching mechanism of the graphene plasmon. When the Fermi level is low and the plasmon energy is greater than $2E_f$, the plasmon is in an interband Landau damping regime and no plasmon effect can be observed(off state). This scenario corresponds to the sketch in the middle of the upper panel. On the other hand, when the Fermi level is high (the other two sketches), the plasmon has no interband Landau damping and near field imaging has strong fringe structures(on state). By tuning the electrical gate, the plamson can be readily switched on/off, which is promising for plasmon transistors. The lower panel of Figure4E illustrates the calculated plasmon amplitude color map as a function of the gate voltage and plasmon wavelength. Plasmon dispersion relation is shown by dashed lines. We can see that the plasmon intensity is barely discernable, if not zero, when the gate voltage $V_B$-$V_D$ is below 5 V. Also shown in the figure are extracted data points for plasmon wavelength from the experiment[21].

The study of graphene plasmon with this near field technique has obtained lots of fascinating results, thanks to its informative character. For instance, Woessner *et al.* identified that the graphene plasmon has a long lifetime when it's sitting on a thin h-BN flake[33]. In addition to graphene plasmonics, the near field technique has also found its arena in phonon polaritons in other 2D materials. The hyperbolicity of h-BN phonon polariton has been intensively investigated[61, 62].

Now let's get back to localized graphene plasmon, which can enhance light absorption in the resonance frequency region. According to Kirchhoff's law, the inverse process of light absorption, i. e., light emission can also be enhanced. Thermal emission spectroscopy has been used to characterize many materials. Recently, Brar *et al.* utilized this technique to study the localized plasmon thermal emission in graphene nano-antenna, mostly composed of graphene nanoribbons[63]. Figure5A is the setup used in the experiment. Figure5B shows the emission spectra of a graphene nanoribbon array at 250 °C, with two carrier concentrations. By comparing the emission spectra of the near-zero doping sample and the sample with certain carrier density, the change of the emissivity can be obtained. Figure5B shows the change of emissivity at carrier density $1.2\times10^{13}$ cm$^{-2}$ and resonance peaks can be observed, which correspond to the plasmon resonances. Again, the emissivity can be tuned by the gate voltage. Figure5C plots the emissivity spectra for different carrier densities. As expected, a higher carrier density sample emits more infrared light. Since the plasmon structures are nanoribbons, they are anisotropic. This has consequences, such as polarized light emission. Indeed, this is exactly the case, as the spectra displayed in Figure5D for different detection polarizations, with 90° giving strongest thermal emission. This experiment opens an avenue for graphene plasmonic structures to be utilized in mid-IR light sources.

## 4  Plasmon-phonon coupling

Atomically thin graphene is vulnerable to the immediate environment. Many factors, such as doping, strain of graphene can be strongly influenced by the substrate. Plasmons in graphene can be influenced to a large extent by the environment as well. For instance, the dielectric constant of the substrate is one of the determining factor for the plasmon resonance frequency[24]. More importantly, a polar substrate, whose surface phonon carries dipole moments, will qualitatively change the plasmon dispersion in graphene, as mentioned in a previous section for graphene ribbons on SiO$_2$/Si substrate. Surprisingly, graphene plasmon can have very strong interaction to other atomically thin 2D materials with polar phonons.

Hexagonal-boron nitride (h-BN) has been widely used in graphene devices due to its ability to enhance the carrier mobility[64]. Brar *et al.*[26] and Jia *et al.*[27] have studied localized graphpene plasmons on atomically thin h-BN surface. Even though the h-BN flake used in the study is atomically thin, the effect on the plasmon is very dramatic. Figure6A illustrates a graphene ribbon on SiO$_2$ substrate, with a single h-BN layer in-between. The charge carrier collective oscillation and the h-BN lattice vibration are artistically sketched. Figure6B shows the extinction spectra of such graphene ribbons with various ribbon widths. In the same figure, the bare h-BN extinction spectrum is shown and the polar phonon at 1370cm$^{-1}$ is salient. Clearly, the graphene plasmon mode hybridizes with the h-BN polar phonon mode and the plasmon dispersion exhibits an anti-crossing behavior. Since the samples are also on SiO$_2$/Si substrate, graphene plasmon also hybridizes with SiO$_2$ surface polar phonons at 800 cm$^{-1}$ and 1100cm$^{-1}$ and four resonance modes can be observed in a single spectrum in most cases. Figure6C presents a loss function map based on RPA calculation. Four plasmon branches and their dispersions are shown. Two polar phonons from SiO$_2$ and one phonon from h-BN are indicated. From the anti-crossing behavior and the splitting of the two relevant branches, it can be inferred that the coupling between graphene plasmon and h-BN phonon is in the classical electromagnetic strong coupling regime. It's an astonishing result, given the fact that the h-BN here is only one atom layer. This is a direct consequence of the strong energy confinement of graphene plasmons, with a mode volume ~$10^7$ times smaller than the free space counterpart[26].

Graphene plasmon not only can couple to external polar phonons, but also couple to intrinsic ones. However, for single layer graphene, there is no intrinsic polar phonon. AB- stacking bilayer graphene, which is typically obtained through mechanical exfoliation of bulk graphite, has a polar phonon and the plasmon in it can interact. Yan *et al.*[53] studied plasmons in bilayer graphene in very detail. Figure7A shows the absorption spectrum of a bilayer graphene without patterning into nano-structures. An IR-active phonon at 1580cm$^{-1}$ can be observed. Meanwhile, a broader peak around 3500cm$^{-1}$ is shown, which is due to the electronic transitions, as indicated in the lower-right inset. The plasmon-phonon coupling strength in bilayer graphene is similar to that of graphene on single layer h-BN. Both qualitative and quantitative behaviors are similar. Figure7B presents the plasmon spectra for a bilayer graphene ribbon array with various chemical doping levels. The ribbon width is designed to be 100nm. The two hybrid plasmon-phonon modes show anti-crossing behavior. At certain point, when the plasmon frequency coincides with the phonon frequency, the spectrum exhibits a broad peak with a sharp dip in the center. This is so called phonon-induced transparency, a classical analog of the EIT (electromagnetically induced transparency)phenomena[65]. This phonon-induced transparency can be tuned either by chemical doping, or electro-static gating, as shown in Figure7B and 7C. Yan *et al.*[53] have demonstrated that AB-stacking bilayer graphene is a very attractive plasmonic material, with potential applications in tunable slow-light devices.

## 5 Potential applications

Graphene shows huge potential in photonic and optoelectronic applications[66]. By combining graphene with metal-based structures, reconfigurable metamaterials[67] and metasurfaces[68] in the terahertz frequency range have been achieved. Localized plasmons in graphene nanostructures are in the mid-IR frequency range and so are biological and organic molecule vibrational frequencies. As a result, graphene plasmon possesses huge potential in molecular sensing. Besides, graphene plasmon can enhance the light absorption and emission efficiency, which can be utilized in photo-detection and light sources.

Li *et al.* are among the first to detect thin layers of organics through graphene plasmon enhanced infrared spectroscopy[24]. Figure8A presents the infrared absorption spectra of a thin layer of PMMA on a graphene nanoribbon array. Both spectra for parallel and perpendicular polarizations are shown. The vibration feature for carbonyl bond at around 1750cm$^{-1}$ is very prominent when the plasmon in graphene is excited. The sharp vibration mode and the relatively broad graphene palsmon mode couple strongly and form a Fano resonance system[69], which has very strong effect on the lineshape of the vibration mode, depending on the relative peak frequencies of the two modes.

Plasmons based on noble metals have been applied to detect biological molecules too[70]. One may ask, what's the novelty and advantage for graphene in this application? In the study by Rodrigo *et al.*[25], this question is answered. First, graphene plasmon offers superior tunability and the sensing frequency range can be tuned in a wide range for a single plasmonic device, which makes it capable of selective sensing. Second, graphene plasmon exhibits very strong energy confinement and the sensitivity for detecting a small amount of molecules is much higher. Figure8B plots the extinction spectra of a protein layer on graphene nanoribbon antenna and gold antenna. The features due to protein on graphene nanoribbon array are much more prominent than those on gold antenna. This is a direct experimental comparison for noble metal and graphene plasmon in biological applications and graphene overshadows noble metals. Due to the stronger energy confinement in graphene antenna array, the protein thin layer is in a much stronger electromagnetic field. Figure8C shows the calculated electrical field spatial distribution using a finite element method for both graphene antenna and gold antenna. Electric field is much more confined in the vicinity of graphene antenna. The hot spots are at the ends of the gold antenna and along the edges of the graphene nanoribbon. Figure8D plots the percentage of the near field intensity confined within a given distance *d* from the structure surface and the inset is the enlarged version in the small *d* regime. We can see that graphene antenna confines 90% of the electric field within 20nm distance from the surface, while gold antenna field intensity spreads out to hundreds of nanometers.

Grahene plasmon has the capability to enhance light detection and emission efficiencies in the resonance frequencies. This has been demonstrated by Tong *et al.*[71]in the terahertz frequency range, and by Freitag *et al.*[72] in the mid-IR range. Freitag *et al.* fabricated a graphene nanoribbon array as the mid-IR photo-detector, as shown in Figure8E for the device configuration. The Fermi level of the ribbon array was tuned by the back gate. Figure8F shows the photo-response as a function of the gate voltage for both light polarizations. With incident light polarization perpendicular to the ribbons, the response is much stronger(10 times), due to the plasmon enhanced light absorption. By varying the gate voltage, a resonance peak shows up which indicates the photon energy is in resonance with graphene plasmon mode and the response is the largest in that case. This is a nice demonstration of polarization sensitive and gate tunable photo-detection based on graphene plasmonic structures.

# 6 Summary and outlook

In this review article, we mainly focus on the plasmons from graphene, such as terahertz plasmons, mid-IR plasmons, plasmon-phonon interaction and some potential applications. Both localized plasmons and propagating plasmons are discussed and two major techniques, absorption spectroscopy and s-SNOM imaging, are introduced. For the fast-moving field of graphene plasmonics, these topics are only a limited portion of the whole achievements in recent years.

Graphene plasmonic response depends on the carrier density. Lightly doped graphene shows weak response. For graphene plasmonic applications, one typically needs doped samples with high carrier density. Chemical doping is a convenient way and it can efficiently enhance the carrier density. However, such doping is usually unstable. The carrier concentration decreases with time. It's desirable to develop stable doping methods, which will facilitate the progress of graphene plasmonics.

Many investigations have been devoted to the interaction of graphene plasmons with phonons. Less efforts have been put on other kinds of interactions, such as plasmon-electron interaction[73] and plasmon-exciton interaction[74]. Graphene plasmon interaction with electrons can form new quasi-particles termed plasmarons[73]. Plamson-exciton interaction can form plexcitons[74], which is a hot topic for metal plasmonics recently. More studies for plasmarons and plexcitons for graphene will be performed in the near future.

Graphene is an excellent plasmonic material. However, to reach its full potential, it's necessary to combine graphene with other materials, such as other two-dimensional materials[75] and noble metals. Meanwhile, the research for graphene metamaterial and metasurface is still in its infancy, more research efforts and more research groups will be participated in this endeavor. Graphene is the first widely studied 2D material and other 2D materials have become popular too[76]. Plasmons in other 2D materials, such as phosphorene[77, 78], will be of great interest for the nanophotonics community.

**Acknowledgements:**

This project is supported by the National Key Research and Development Program of

China(Grant number: 2016YFA0203900) and Oriental Scholar Program from Shanghai Municipal Education Commission.

Figure captions

**Figure 1** Terahertz plasmons of graphene micro-structures. (A) AFM image of a graphene micro-ribbon array with ribbon width $W$=4μm. (B) Extinction spectra for micro-ribbon arrays with ribbon width $W$=4,2,1 μm respectively. The intensities are normalized to have the same peak height. (C) Normalized plasmon frequency as a function of Fermi energy and carrier density. The plasmon frequencies are normalized with the ribbon width. Therefore, the data points for 1,2,4 μm ribbons are collapsed to the same square root scaling relation (solid curve). The dashed line is the best fit using a linear dependence, which deviates from the data. (D) A SEM image of a graphene micro-disk array. (E) Extinction spectra for 1-layer, 2-layer and 5-layer disks. (F) Normalized resonance frequency as a function of the normalized carrier density for multilayer structures(grey squares) and a single layer case(red squares)( A–C are reprinted with permission from[14] , Copyright 2011 Nature Publishing Group. D-F are reprinted with permission from[15]. Copyright 2012 Nature Publishing Group).

**Figure 2** Magnetoplasmons in graphene. (A) Edge and bulk plasmon modes for a graphene micro-disk in a magnetic field. (B) Extinction spectrum for a graphene disk array at B=16T. (C) The FWHMs for the edge mode and bulk mode as a function of the magnetic field. Solid curves are fittings described in[28]. (D)Illustration of exciting edge plasmon using an electrical pulse. (E) Magnetic field dependence of the carrier drift velocity around the edge(A-C are reprinted with permission from [28], Copyright 2012 American Chemical Society; D-E are reprinted with permission from[31], Copyright 2013 Nature Publishing Group).

**Figure 3** Mid-IR localized plasmons in graphene. (A)SEM image of a graphene nanoribbon array. (B)Typical extinction spectra for graphene nanoribbons on $SiO_2$/Si substrate, with both polarizations. (C) Calculated loss function as a function of wave-vector and frequency. Surface polar phonons and graphene intrinsic phonon are indicated. (D)Extinction spectra for a graphene nanoribbon array with frequency in the vicinity of the optical phonon frequency. The inset shows the FWHM as a function of the resonance frequency. (E) An illustration for the phonon-assisted plasmon decay channel(A-E are reproduced with permission from [16], Copyright 2013 Nature Publishing Group).

**Figure 4** Propagating mid-IR plasmons in graphene. (A) Illustration of the s-SNOM experimental configuration. (B) A typical s-SNOM imaging of graphene plasmon near graphene edges. The graphene sample is on the $SiO_2$/Si substrate. (C) Gate dependence of the imaging fringe amplitude. The inset shows the gate-tunable device arrangement. (D) Gate dependence of the plasmon wavelength. The gate-dependent optical conductivity (real and imaginary parts) is shown in the inset. (E) Lower panel: plasmon wavelength(data points) obtained from experiment as a function of the gate voltage. Dashed line is the calculated result. The plasmon intensity is indicated by the color map. Upper panel: Illustration of the plasmon inerband Landau damping. Red cross indicates that the damping cannot happen and the plasmon has strong amplitude(A-D are reprinted with permission from[20],E is reproduced with permission from [21], Copyright 2013 Nature Publishing Group).

**Figure 5** Plasmonic thermal emission in graphene. (A) Experimental setup for the thermal emission studies. (B) Thermal emission spectra of a ribbon array at 250 °C at charge neutral and heavily doped regimes. (C)Doping dependence of the emissivity enhancement spectrum. (D)Polarization dependence of the emissivity enhancement spectrum(A-D are reproduced with permission from [63], Copyright 2015 Nature Publishing Group ).

Figure 6 Plasmon-phonon coupling for graphene plasmons. (A) A sketch for graphene plasmon and h-BN polar phonons. (B) Extinction spectra of graphene nanoribbons with different ribbon widths on single layer h-BN, sitting on SiO2/Si substrate. A bare h-BN spectrum is also shown in the figure. (C) Calculated loss function map using RPA method. Extracted plasmon frequencies are also shown in the figure as discrete points(A-C are reproduced with permission from[26], Copyright 2014 American Chemical Society).

Figure 7 Plasmons in AB-stacking bilayer graphene. (A) Extinction spectrum of unpatterned bilayer graphene. Upper inset shows the infrared active phonon mode and the other inset illustrates the electronic transitions. (B) Extinction spectra for a bilayer graphene nanoribbon array with different chemical doping levels. (C) Gate-tuning of the bilayer graphene plasmon spectrum(A-C are reprinted with permission from[53], Copyright 2014 American Chemical Society).

Figure 8 Graphene plasmon applications in sensing and photo-detection. (A) Infrared absorption spectrum of graphene nanoribbons covered with a thin layer of PMMA. (B) Extinction spectra of graphene and gold biosensors covered with a protein layer. The thick curves are the measured spectra and the thin curves are the fittings for bare graphene and gold biosensors. (C)

Calculated near field distribution for gold(upper panel) and graphene(lower panel) biosensors at frequency 1600cm$^{-1}$. (D) Field confinement percentage as a function of the distance from the biosensor surface for both graphene and gold. The inset is the enlarged version in the small distance *d* regime. (E) A graphene nanoribbon photodetector with electrical gate. (F) Gate and polarization dependence of the photo-response for the ribbon array shown in (E). The red dots are for the perpendicular polarization, while the blue dots are for the parallel case(A is reproduced by permission from [24], Copyright 2014 American Chemical Society. B-D are reproduced by permission from [25],Copyright 2015 AAAS. E-F are reprinted by permission from[72], Copyright 2013 Nature Publishing Group).

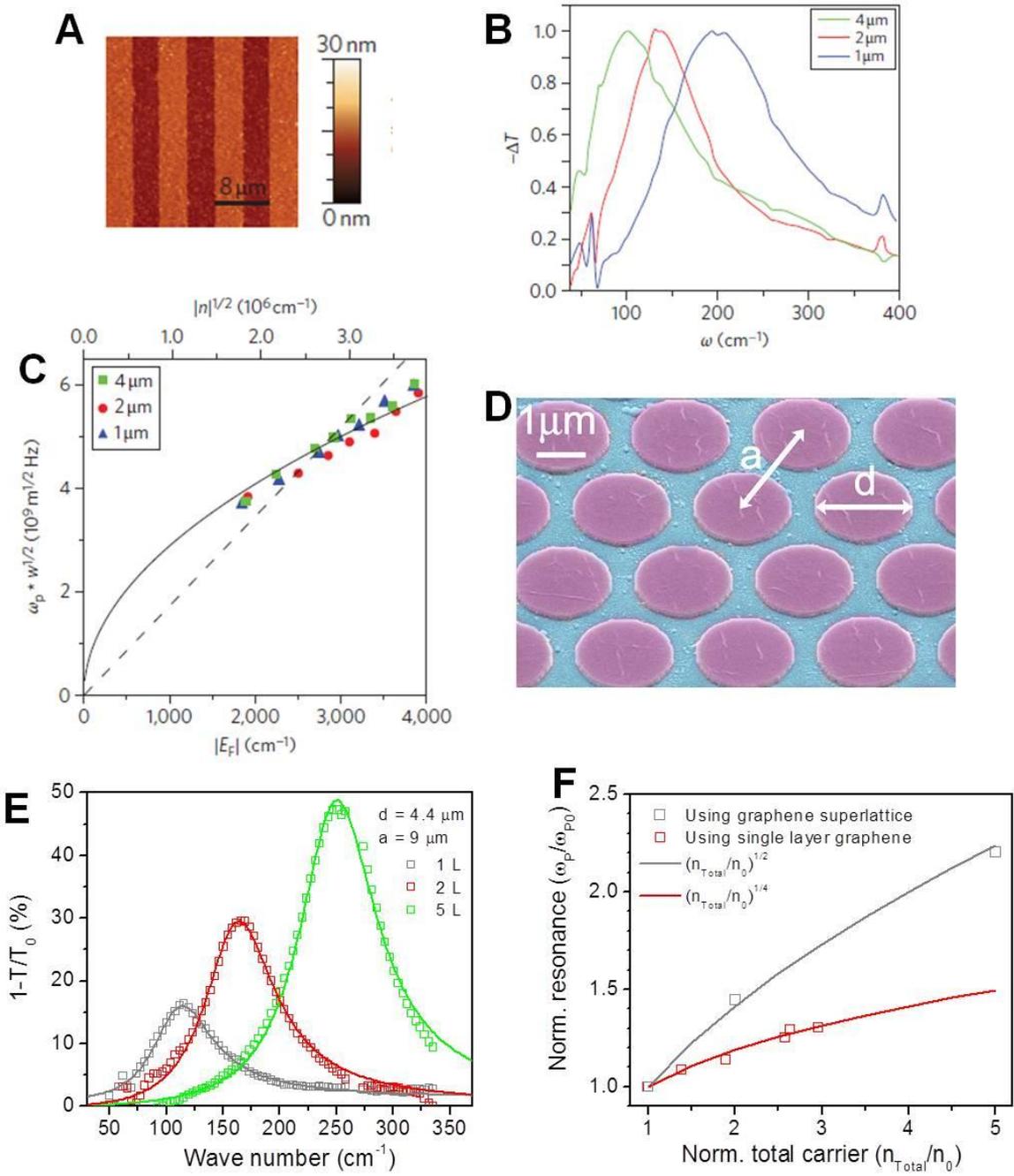

Figure 2

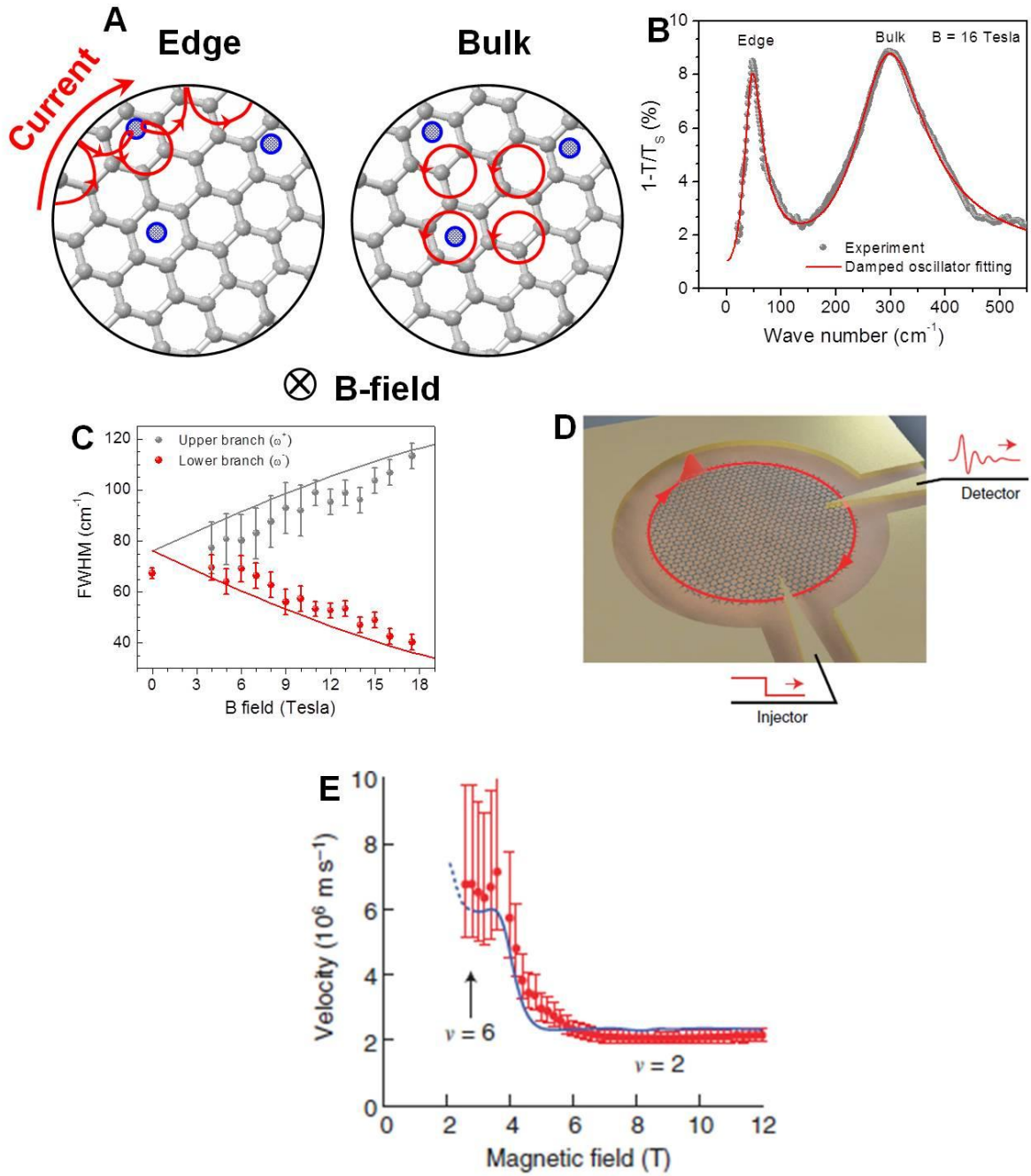

Figure 3

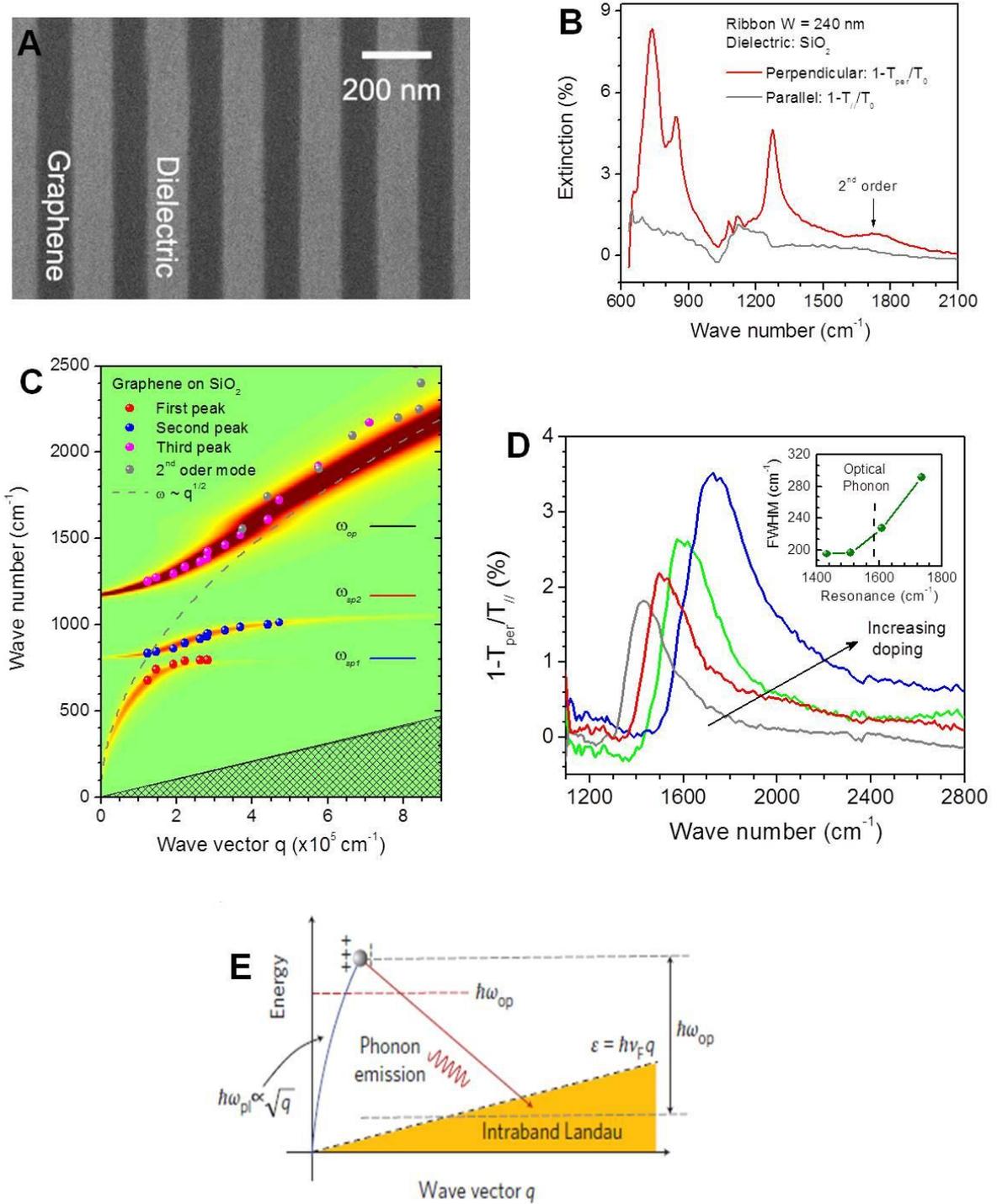

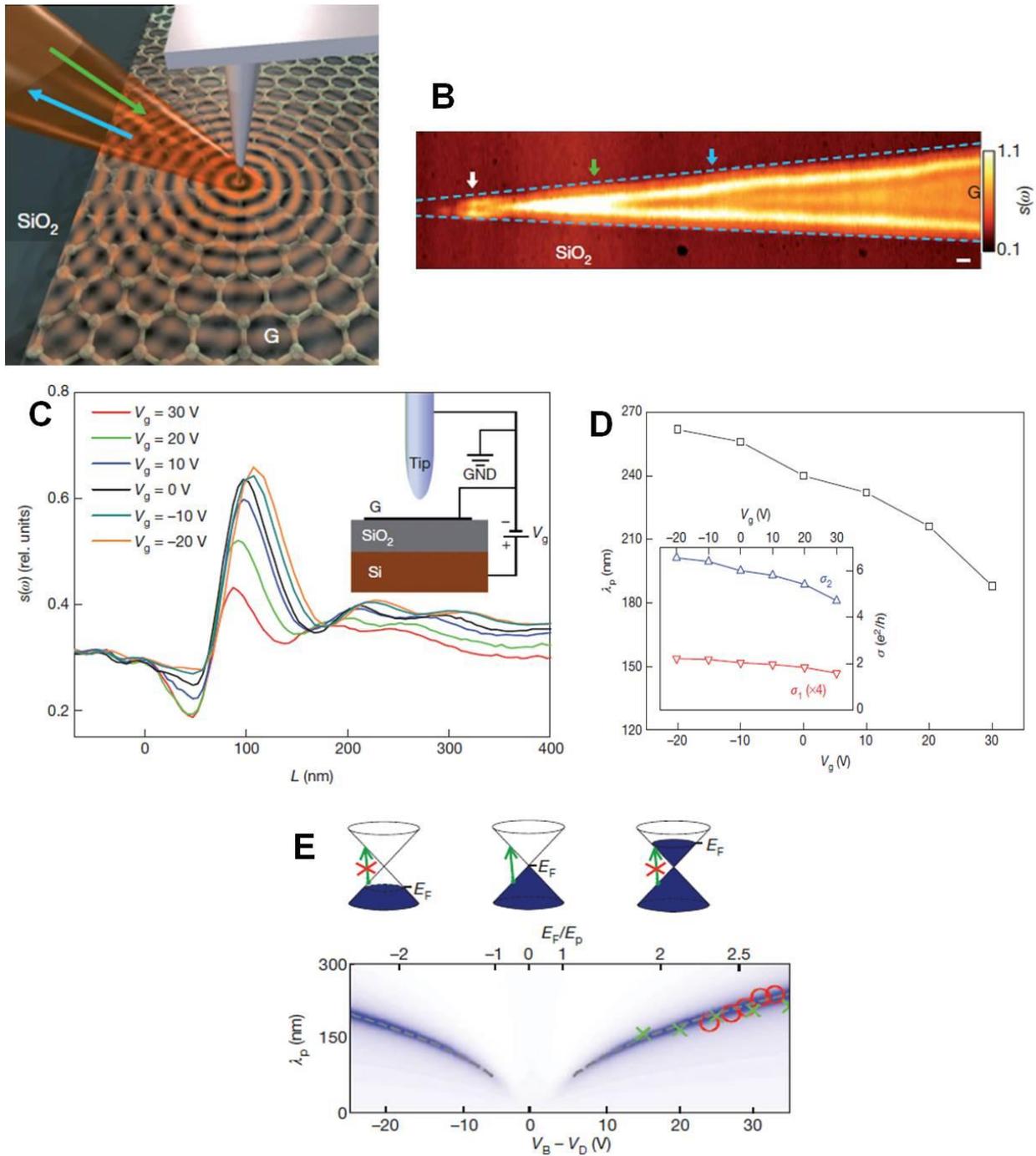

Figure 5

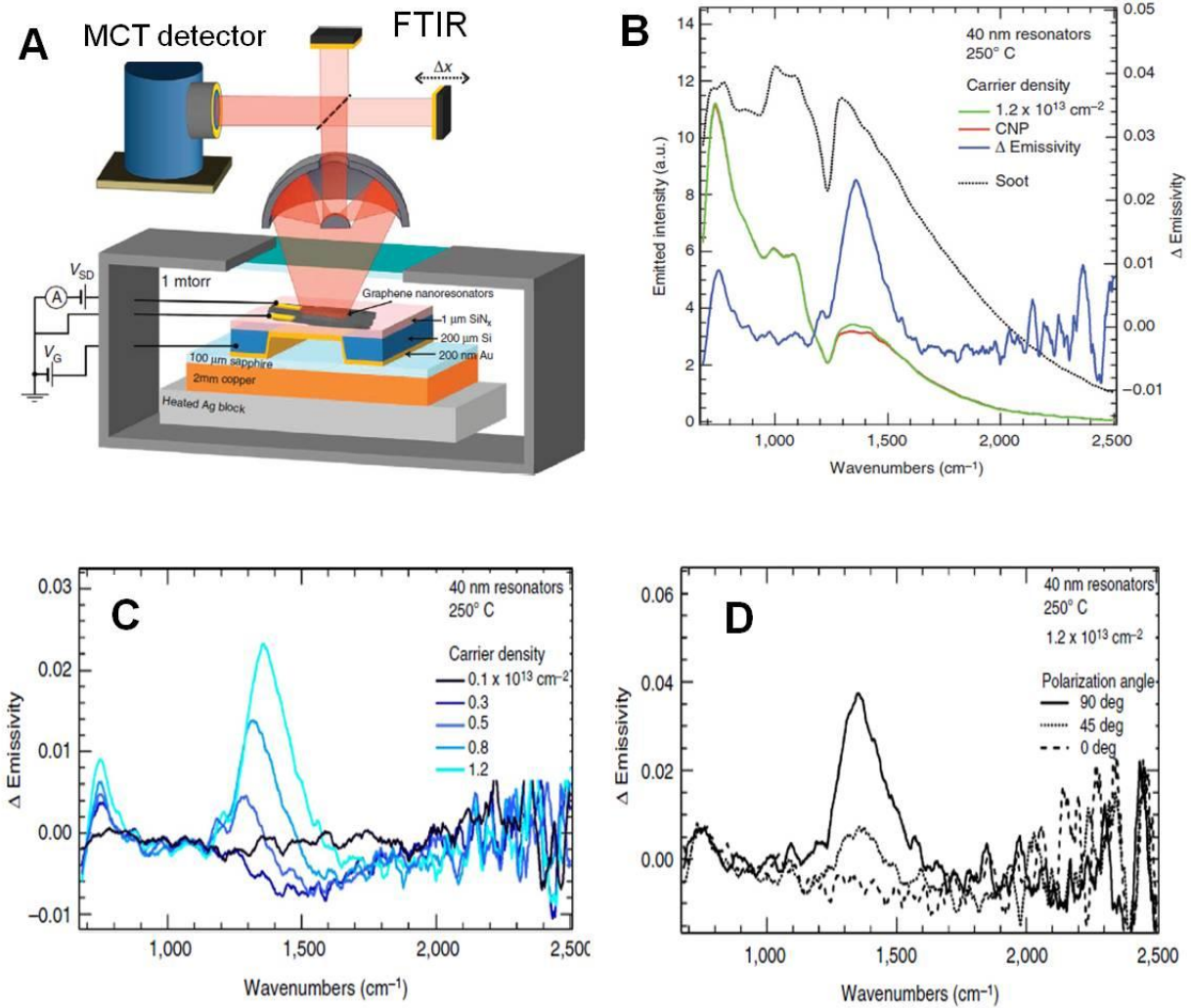

Figure 6

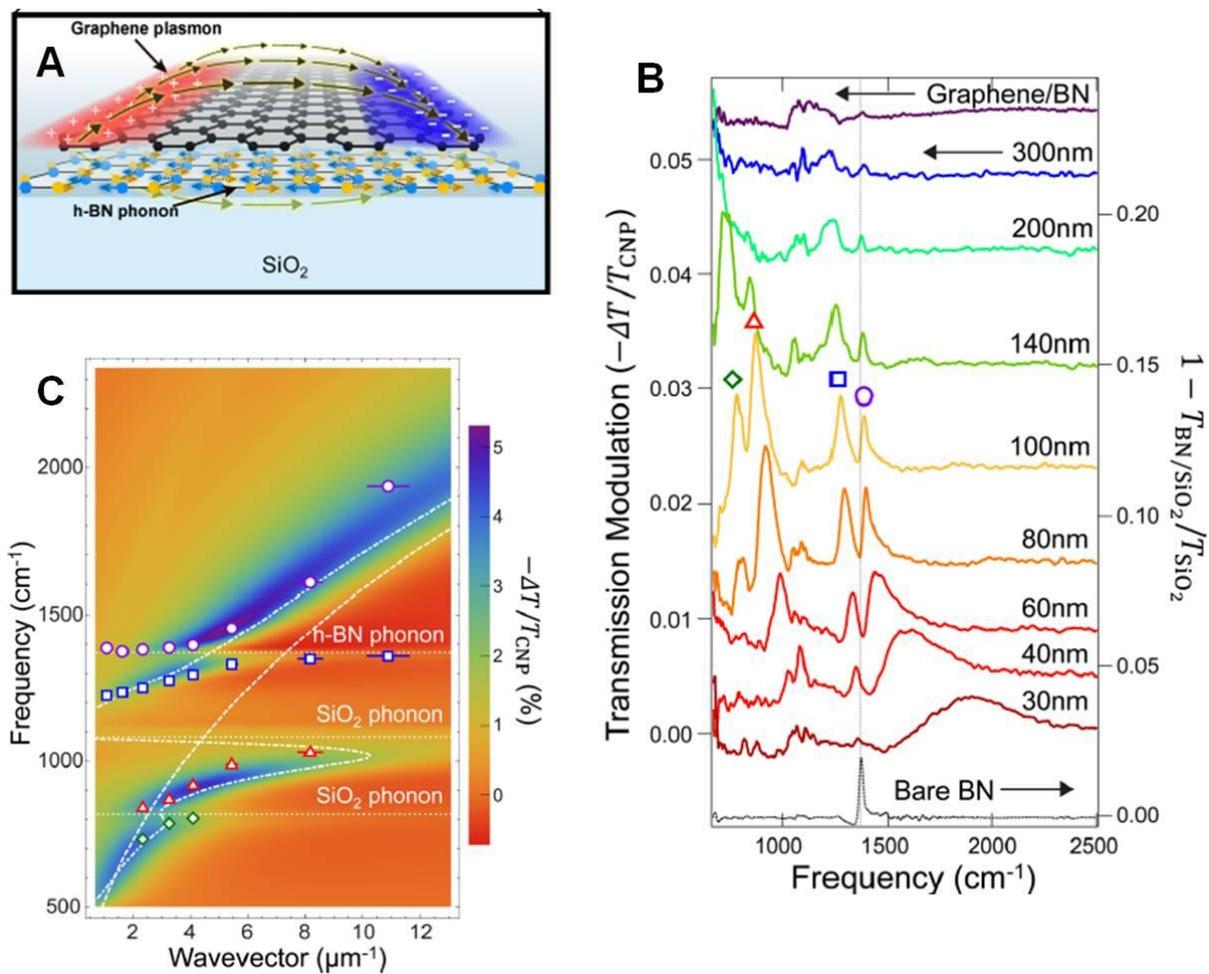

Figure 7

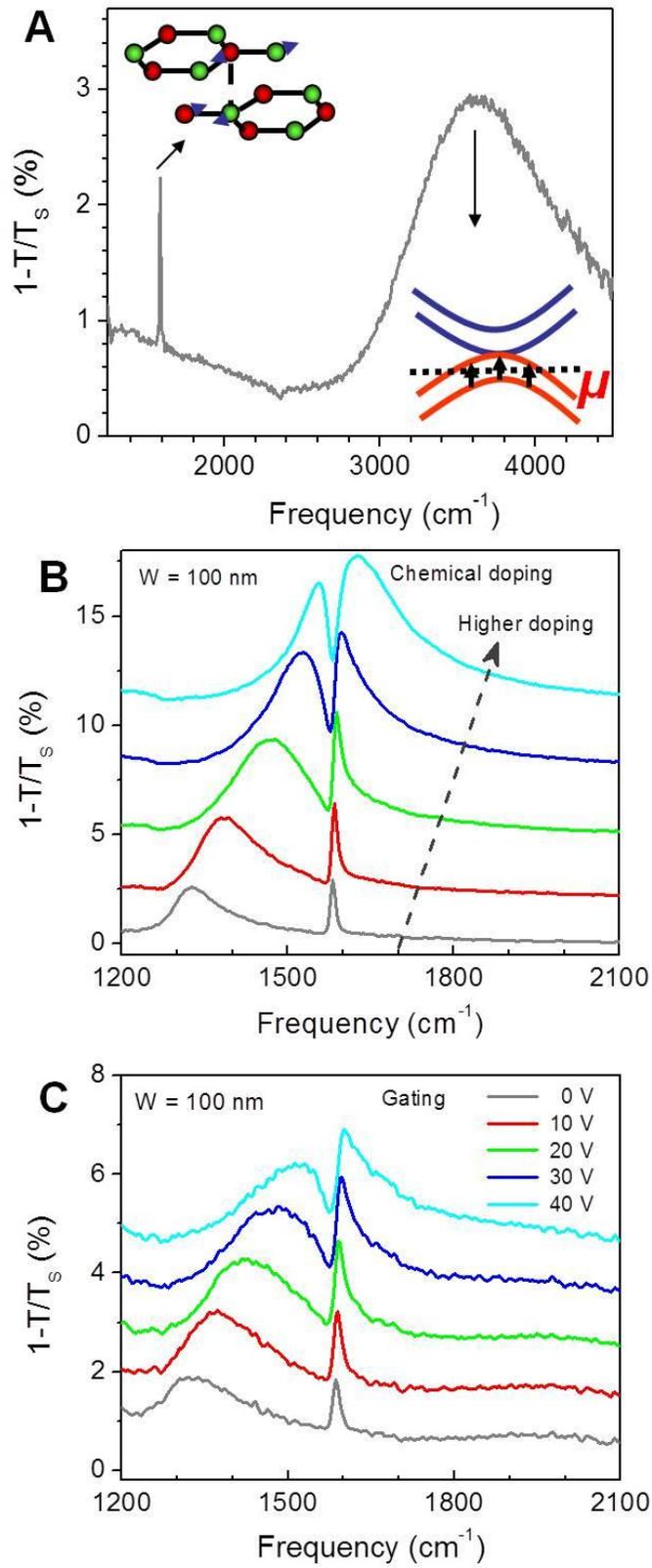

Figure 8

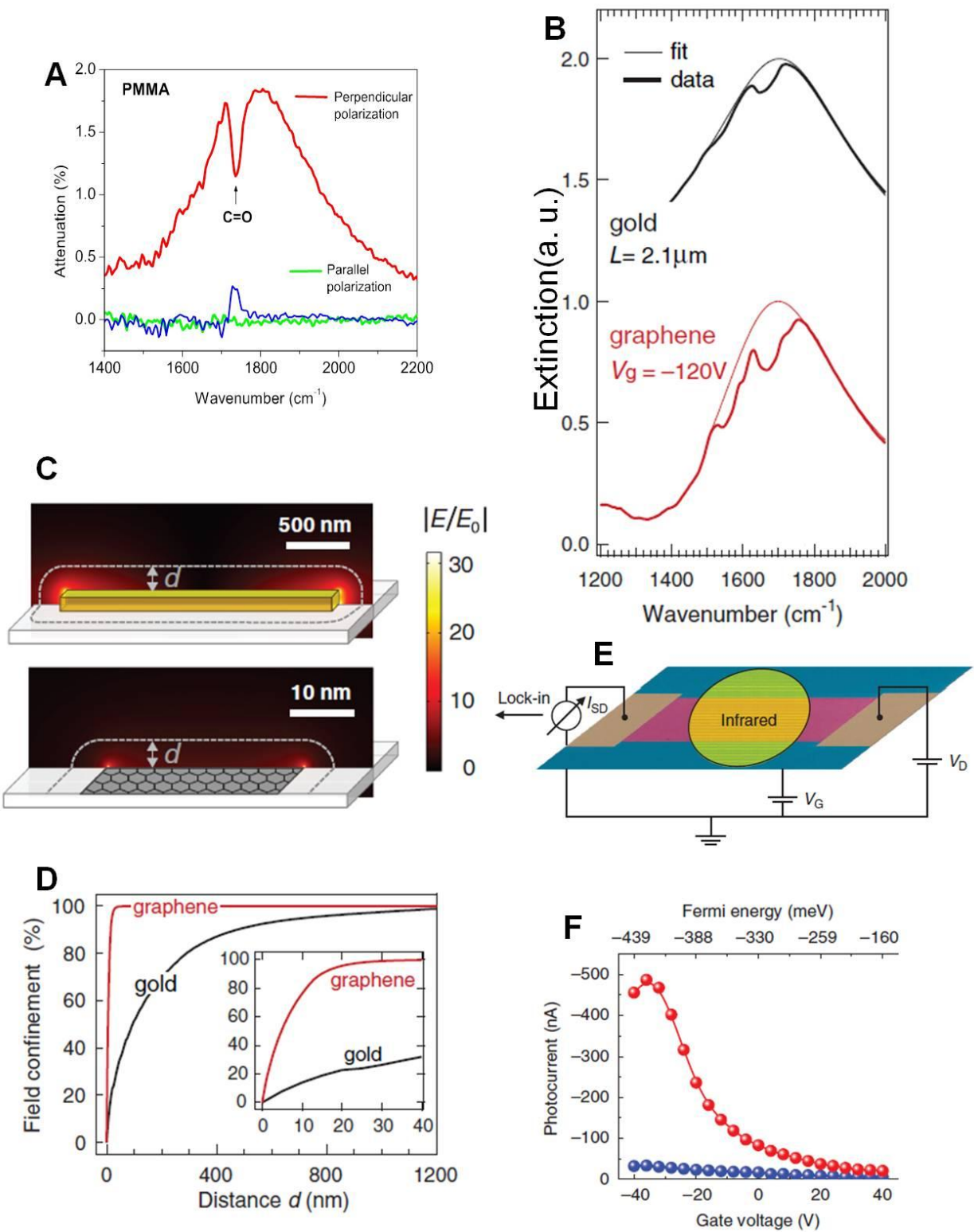